\documentstyle[avpa,epsfig]{qcdparis}
\begin{document}
\newcommand{\beq}{\begin{equation}}
\newcommand{\eeq}{\end{equation}}
\newcommand{\beqa}{\begin{eqnarray}}
\newcommand{\eeqa}{\end{eqnarray}}
\newcommand{\gsim}{\raisebox{-0.07cm}{$\stackrel{>}{{\scriptstyle
 \sim}}$} }
\newcommand{\lesim}{\raisebox{-0.07cm}{$\stackrel{<}{{\scriptstyle
 \sim}}$} }
\addtolength{\oddsidemargin}{0.3cm}
\addtolength{\evensidemargin}{0.3cm}
\addtolength{\topmargin}{2.0cm}
\addtolength{\footnotesep}{0.2cm}
\pagestyle{plain}
\thispagestyle{headings}

\title{On dynamical parton distributions of hadrons and photons$
       ^{\,\ast}$}

\author{Andreas Vogt$^{\,\sharp}$}

\affil{Deutsches Elektronen-Synchrotron DESY \\
 Notkestra{\ss}e 85, D-22603 Hamburg, Germany}

\abstract
{The basic concepts and predictions of the `dynamical' GRV parton
distributions for hadrons and photons are discussed. Comparisons of
these predictions with recent experimental results, especially from
HERA, are presented.}
\twocolumn[\maketitle]
\fnm{7}{Talk given at the Workshop on Deep Inelastic Scattering and
QCD, Paris, April 1995.}
\fnm{8}{On leave on absence from Sektion Physik, Universit\"at
M\"unchen, D-80333 Munich, Germany}

\section{Introduction}
In the conventional approach to the proton's parton structure, the
perturbatively uncalculable quark and gluon input densities for the
Altarelli-Parisi (AP) $Q^2$-evolution are fitted, at some resolution
scale $Q_0$ well in the perturbative region, $ Q_0^2 = 2\ldots 5\mbox
{ GeV}^2 $, to an exhaustive set of DIS and related data. See \cite
{MRS,CTEQ} for recent examples. This procedure is sufficient to fix the
parton distributions at all perturbatively accessible scales for $ x\!
>\! x_{min} $, where $x_{min}$ ($>$ 0.01 for pre-HERA DIS data) denotes
the minimal momentum fraction for which enough experimental results are
available to constrain the fit. However, extrapolations to smaller $x$
are notoriously unreliable \cite{MRS,GRV90} in this framework. Hence
this approach does not possess predictive power for the small-$x$
region now accessible at HERA.

One can try to get more out of the AP formalism by assuming a wider
range of validity towards small $Q^2$ for this perturbative twist-2
renormalization group evolution than in the conservative approach
summarized above. A basic observation in this context is that the
constraint imposed by the energy-momentum sum rule for the parton
distributions becomes more important at smaller $Q^2$. The well-known
maximal example is the purely dynamical generation of the gluon ($g$)
and sea quark ($\bar{q}$) densities \cite{Dynam}. There it is assumed
that the AP equations apply down to a very low scale $\mu_D$, where
the valence quark densities (which evolve separately and are
constrained by the charge sum rules at small $x$) saturate the momentum
sum rule and thus leave no room for non-vanishing sea and gluon inputs.
Hence this approach completely predicts $ g,\, \bar{q}\, (x,Q^2 \! >\!
\mu_D^2) $ dynamically at all $x$. These predictions, however, are too
steep in $x$ as compared to experimental constraints \cite{GRV90}.
Theoretical objections have also been raised \cite{BrSc}.

In this talk we will give a brief summary of a less ambitious, but
theoretically more sound approach, in which all parton densities
are generated from intrinsic valence-like initial distributions
at some low scale $\mu $
determined from large-$x$ proton structure constraints. Usually the
resulting distributions are (somewhat simplifying) also described as
`dynamical'. In fact, this procedure leads to rather unambiguous
dynamical predictions for $g$ and $\bar{q}$ at small $x$ \cite
{GRVp,GRV93,GRV94}. Moreover, the application of this concept to pions
and photons allows for (approximate) predictions of $g^{\pi}$ and
$g^{\gamma}$ also at large $x$ \cite{GRVpi,GRVph}. We will focus on
these phenomenological predictions and their comparison to recent
experimental results, most noteably from HERA. See \cite{GRV94} for a
discussion of theoretical issues related to this approach.
\section{Partons in the proton}
One place where the purely dynamical predictions are too small is
fixed-target direct-photon (DP) production, $ pp \!\rightarrow \!
\gamma X $. This process probes the large-$x$ gluon density, $x$ \gsim
0.3.
Thus the DP data enforce a
substantially harder gluon distribution \cite{GRV90}. Together with the
momentum sum rule constraint, these results lead to a valence-like form
of $g(x,\mu^2)$, $ g \sim u_v $, if a sufficiently small input scale
$\mu $ is chosen. Since evolution from a sizeably smaller scale would
need a physically unreasonable input ($g$ harder than $u_v$), $\mu $
can be assumed to be the lowest scale down to which the perturbative
AP equations may hold. The basic assumption is that $\mu $ can actually
be reached perturbatively.

The most simple valence-like gluon and sea quark input ansatz has been
employed to fix $\mu $ \cite{GRVp}:
\beq
   xg(x,\mu^2)       = A x^a (1-x)^b           \:\:, \:\:
   x\bar{q}(x,\mu^2) = A' x^{a'} (1-x)^{b'}    \:\: .
\eeq
The free parameters of (1) have been determined from fixed-target DIS
and DP data, with the valence distributions $u_v$ and $d_v$ and the QCD
scale parameter $ \Lambda^{(4)} = 0.2 \mbox{ GeV} $ taken from a
conventional fit. This procedure results in $ g \sim u_v $ for $\mu
\simeq 0.55 $ GeV in next-to-leading order (NLO) perturbative QCD,
corresponding to $\alpha_{s}(\mu^2)/\pi \simeq 0.2 $ \cite{GRVp}.
Obviously, the accuracy of this $\mu $ determination is limited by the
accuracy of the large-$x$ data used. An uncertainty of about 10\% has
been estimated in \cite{GRV93}. The complete set of initial
distributions is shown in fig.~1, together with that one of the recent
update \cite{GRV94} incorporating later large-$x$ constraints. In this
update, $ \mu = 0.58 $ GeV.
\begin{figure}[b]
\begin{center}
\vspace*{-0.3cm}
\epsfig{file=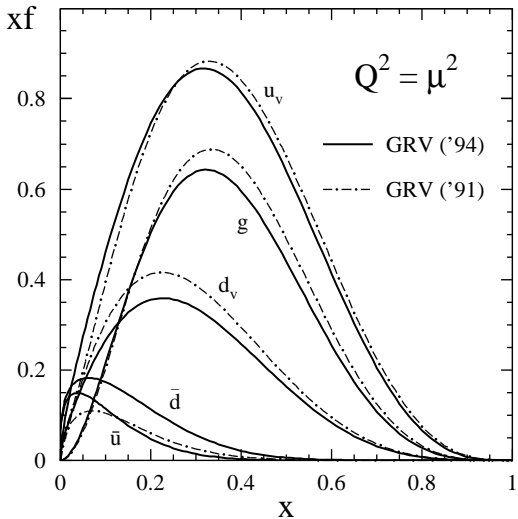,height=7.0cm}
\vspace*{-0.9cm}
\end{center}
\caption[]
 {The NLO valence-like GRV input distributions. $ \mu^2 $ = 0.30
  (0.34) GeV$^2$ for the '91 ('94) parametrizations \ \protect\cite
  {GRVp,GRV94}. \ The strange sea $ s = \bar {s} $ vanishes at $ Q^2 =
  \mu^2 $. $ \bar{u} = \bar{d} $ in GRV ('91).}
\end{figure}

At large-$x$, $ x\! >\! 10^{-2} $, this approach comes close to the
conventional procedure if only the partons in the proton are considered.
At small-$x$ and $ Q^2 \gg \mu^2 $, however, the behaviour of the gluon
and sea quark densities is due to the QCD dynamics here, since
ambiguities from the initial distributions are suppressed by their
valence-like form (small at small-$x$) and the long evolution distance.
Hence the predictive power of the purely dynamical approach \cite
{Dynam} is retained in this regime. The uncertainty of $\mu $ mentioned
above leads to rather moderate an uncertainty of the small-$x$
predictions amounting to, e.g., less than 20\% (10\%) at $ x = 10^{-4}\,
(10^{-3}) $ for $ Q^2 \gg \mu^2 $ \cite{GRV93}. The dependence on the
precise value of $\Lambda^{(4)}$ is also small, see \cite{av}.

\begin{figure}[b]
\begin{center}
\vspace*{-0.4cm}
\epsfig{file=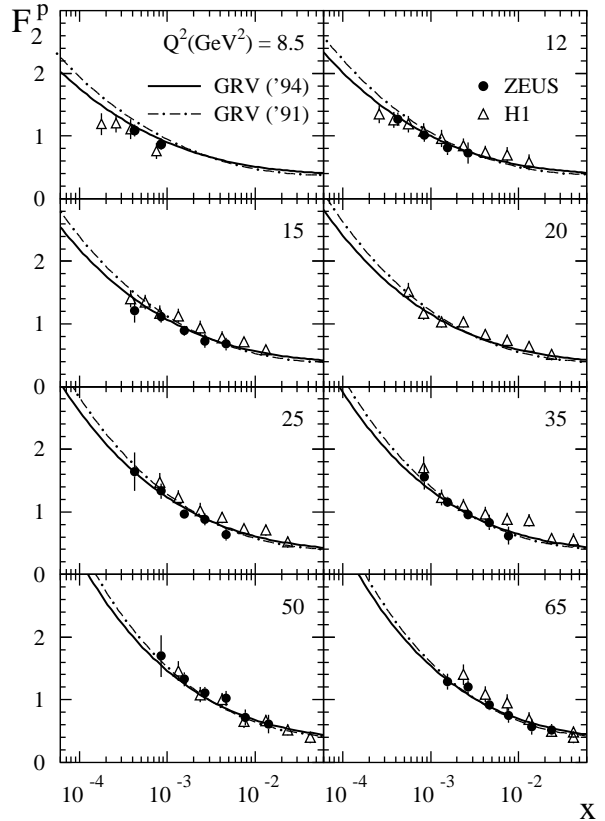,height=11.0cm}
\vspace*{-0.4cm}
\end{center}
\caption[]
 {\ NLO dynamical small-$x$ GRV predictions \cite{GRVp,GRV94} for
  $F_2^{\, p}$ vs.\ recent HERA data \protect\cite{ZEf2,H1f2}. The
  charm contribution has been calculated using the NLO massive
  expressions of \protect\cite{SRvN}.}
\end{figure}
\begin{figure}[bt]
\begin{center}
\vspace*{-0.3cm}
\epsfig{file=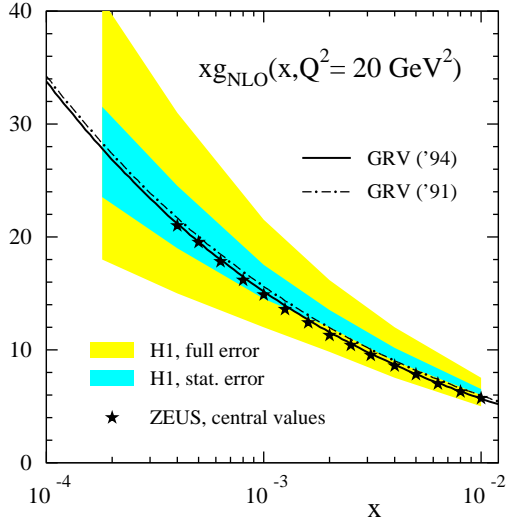,height=7.0cm}
\vspace*{-0.9cm}
\end{center}
\caption[]
 {The dynamical predictions for the NLO gluon density at small-$x$
 \protect\cite{GRVp,GRV94} as compared with constraints derived from
 $\, F_2^{\, p}$ scaling violations at HERA \protect\cite{ZEgl,H1gl}.}
\end{figure}
\begin{figure}[bt]
\begin{center}
\vspace*{-0.3cm}
\epsfig{file=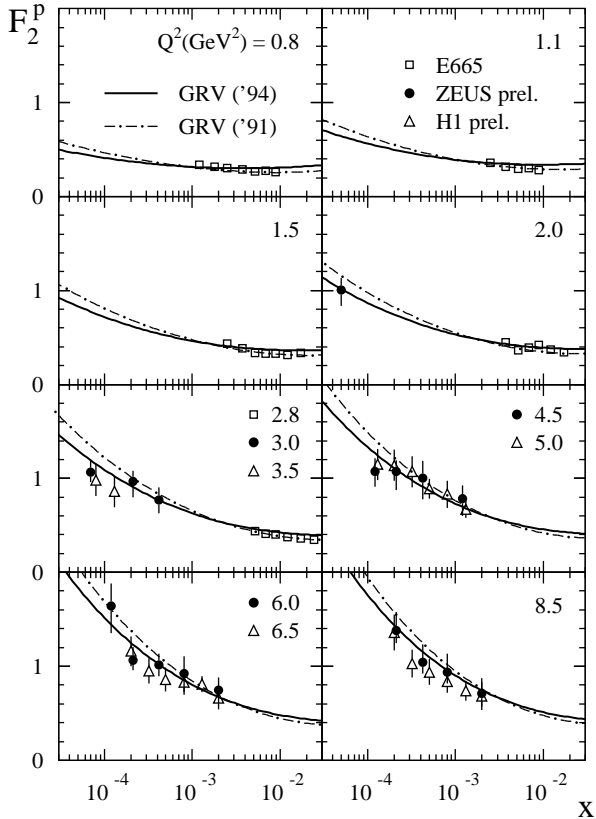,height=11.0cm}
\vspace*{-0.4cm}
\end{center}
\caption[]
 {\ As fig.~2, but with the preliminary low-$Q^2$ HERA results
  \protect\cite{ZEf2p,H1f2p} and the final E665 data \protect\cite
  {E665}. At $ Q^2 \! >\! 2 \mbox{ GeV}^2 $, the curves have
  been calculated for the ZEUS $Q^2$-values.}
\end{figure}
These dynamical small-$x$ predictions \cite{GRVp,GRV94} are compared
with recently published HERA results on the structure function
$F_2^{\, p}$ (rather directly measuring the quark distributions) and
on the gluon density (constrained by the observed scaling violations in
$F_2^{\, p}$) in fig.~2 and fig.~3, respectively. Very recently, the
HERA collaborations have extended their $F_2$ determination to lower
$Q^2$ \cite{ZEf2p,H1f2p}, and the final small-$Q^2$ E665 data have been
presented \cite{E665}. In fig.~4 the GRV predictions are also confronted
to these results. Within their uncertainty indicated above, these
(leading-twist) predictions quantitatively describe all present
structure function measurements at small-$x$ down to $ x\! <\! 10^{-4}
$, including the transition from a rather flat behaviour at $Q^2 \!
<\! 1 \mbox{ GeV}^2$ to the steeply rising $F_2$ at high $Q^2$.
\section{Partons in the pion}
The parton distributions of the pion are experimentally much less
constrained than those of the nucleon. Information on the valence
density $v^{\pi}$ (at $x$ \gsim 0.2) and on the the gluon distribution
$g^{\pi}$ (at $x$ \gsim 0.3) has been inferred from pionic lepton-pair
and direct-photon production, respectively, see \cite{ABFKW,SMRS}.
Virtually nothing is known about the pionic sea quark density
$\xi^{\pi}$.

Nevertheless, the available constraints allow for an additional test
of the concept of valence-like low-scale input distributions. As
discussed above, $\mu $ in (1) represents the lowest scale down to
which the perturbative AP evolution is supposed to hold. Hence $\mu $
should not depend on the hadron under consideration and is taken over
from the proton analysis. In view of the experimental situation
summarized above, the most simple valence-like ansatz is appropriate
here \cite{GRVpi}:
\beq
\label{piin}
    g^{\pi}(x,\mu^{2}) = k \, v^{\pi}(x,\mu^{2}) \:\: , \:\:
  \xi^{\pi}(x,\mu^{2}) = 0  \:\: .
\eeq
Since $k$ is known from the momentum sum rule, (2) provides a
parameter-free prediction of $ g^{\pi}(x,Q^2>\mu^2) $ once
the valence distribution is fixed. In fig.~5 this prediction, using
$v^{\pi}$ of \cite{ABFKW}, is compared to the DP constraints
and perfect agreement is found.
\begin{figure}[hb]
\begin{center}
\vspace*{-0.3cm}
\epsfig{file=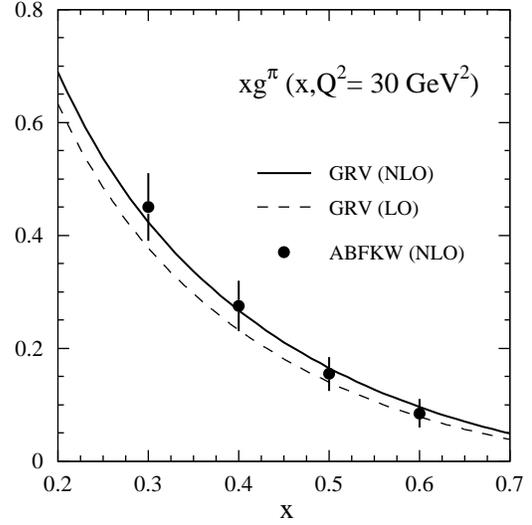,height=7.0cm}
\vspace*{-0.8cm}
\end{center}
\caption[]
 {\ The NLO and LO GRV predictions for the gluon density of the
 pion \protect\cite{GRVpi} in comparison with the NLO error band
 obtained from $ \pi p \rightarrow \gamma X $ (ABFKW) \protect\cite
 {ABFKW}.}
\end{figure}
\section{Partons in the photon}
The quark content $q^{\gamma}$ of the photon has been measured (with
so far rather limited precision) at $e^+ e^-$ colliders via the photon
structure function $F_{2}^{\, \gamma}$ for $x$ \gsim 0.05. The gluon
density $g^{\gamma}$ is virtually unconstrained by these data, see
\cite{av1}. A usual assumption is that at some low resolution scale the
partonic structure of the photon is very closely related to those of
the vector mesons (vector meson dominance, VMD). However, it is well
known that for input scales $ Q_0^2 \geq 1 \mbox{ GeV}^2 $ a pure VMD
initial condition for the photon's modified AP equations is not
sufficient to describe the $F_{2}^{\, \gamma}$ data at larger $Q^2$.

In the present approach, it is very natural to impose a pure VMD input
at the scale $\mu $ \cite{GRVph}:
\beq
 (q,g)^{\gamma}(x,\mu^2) = \kappa \frac{4\pi \alpha}{f_{\rho}^{2}}
 (q,g)^{\pi^0}(x,\mu^2)     \:\: .
\eeq
Here $ f_{\rho}^{2}/(4\pi) = 2.2 $ is the $\rho \gamma $-coupling and
the (GRV) parton densities of the pion have been used instead of the
experimentally unknown $\rho $ distributions. $\kappa $ accounts in
the most simple way for the higher-mass vector mesons, one expects
1 \lesim $\kappa $ \lesim 2. In NLO, (3) has been implemented in the
DIS$_{\gamma}$ factorization scheme, since in the photon case the
$\overline{\mbox{MS}} $ scheme is not suited for physically motivated
input shapes \cite{GRVfa}. $\kappa $ has been determined from the
available $F_{2}^{\, \gamma}$ data, resulting in a good fit for
$ \kappa = 1.6 $ in NLO \cite{GRVph}.

Therefore, a pure VMD input at $\mu $ is in fact successful, and
one obtains an approximate VMD-based prediction of $g^{\gamma}$ at
large-$x$ in addition to the typical dynamical small-$x$ predictions.
Besides from the treatment of the higher-mass vector mesons and from
the $ \rho \rightarrow \pi $ substitution in (3), uncertainties of
these predictions arise from the uncertainty of the pion's quark
content: $\xi^{\pi}$ is unknown, and the normalization of $v^{\pi}$ is
uncertain by about 20\% even in the large-$x$ region \cite{ABFKW,SMRS}.
A larger $v^{\pi}$ (as in \cite{SMRS}) would obviously lead, via the
fit of $\kappa $ in (3), to a corresponding decrease of $g^\gamma $.
Fig.\ 6 shows the agreement of this approximate prediction with a first
HERA extraction of $g^\gamma $.
\begin{figure}[b]
\begin{center}
\vspace*{-0.4cm}
\epsfig{file=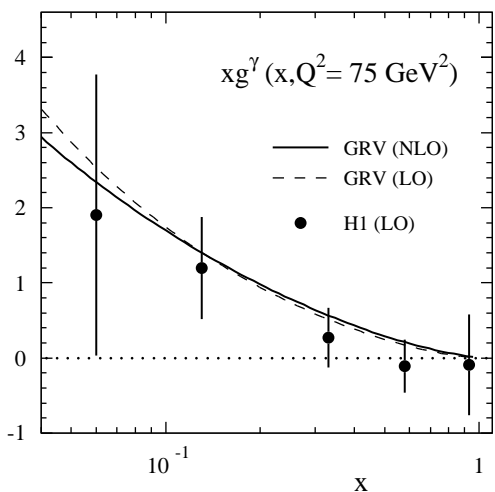,height=6.6cm}
\vspace*{-0.8cm}
\end{center}
\caption[]
 {\ The NLO and LO GRV prediction for the photon's gluon density
  \protect\cite{GRVph} compared with HERA (H1) results from a LO
  analysis of 2-jet photoproduction data \protect\cite{H1glg}.}
\end{figure}
\section{Conclusions}
Valence-like initial distributions at a low resolution scale $\mu $
\lesim 0.6 GeV for $ \Lambda^{(4)} $ = 0.2 GeV allow for a coherent
description of the parton structure of hadrons and photons. Although
this approach is not predictive at large-$x$ in the proton case, it
leads to an essentially parameter-free prediction of the pion's gluon
density in this $x$-region and an (approximate) VMD-based prediction
for the gluon content of the photon. The approach exhibits its full
predictive power at small-$x$, where the gluon and sea quark densities
are generated dynamically. Present data are in very good agreement
with all these predictions.

\vspace{0.5cm}
\noindent
{\bf Acknowledgements}
\vspace{0.3cm}

\noindent
It is a pleasure to thank M. Gl\"uck and E. Reya for a fruitful
collaboration. I am grateful to S. Riemersma for making available the
NLO $F_{2}^{\, {\rm charm }}$ computer code of \cite{SRvN}. This work
was supported by the German Federal Ministry for Research and
Technology under contract No.\ 05 6MU93P.

\end{document}